\documentclass[aps,12pt,final,notitlepage,oneside,onecolumn,%
nobibnotes,nofootinbib,noshowpacs,%
centertags,noshowkeys,amssymb]{revtex4}

\usepackage{graphicx}
\usepackage{amsmath,amsfonts}

\DeclareMathOperator{\Li}{Li}

\newcommand{\nn}{\nonumber\\}
\newcommand{\ii}{\mathrm i}
\newcommand{\dd}{\mathrm d}
\newcommand{\be}{\begin{equation}}
\newcommand{\ee}{\end{equation}}
\newcommand\ba{\begin{eqnarray}}
\newcommand\ea{\end{eqnarray}}

\def\Li#1#2{{\mathrm{Li}}_{#1}\left(#2\right)}

\begin{document}

\title{Radiative corrections to chiral amplitudes in
quasi-peripheral kinematics.}

\author{V. ~Bytev}
\author{E.~Barto\v s}
\altaffiliation{Institute of Physics SAS, 84511 Bratislava,
Slovakia}
\author{M.~V.~Galynskii}
\altaffiliation{Institute of Physics BAS, Minsk, 220072, Belarus}
\author{E.~A.~Kuraev}
\affiliation{Joint Institute of Nuclear Research, 141980 Dubna,
Russia}


\begin{abstract}
Chiral amplitudes for two jets processes in quasi-peripheral kinematics are calculated at
the Born and one loop correction level. The amplitudes of subprocesses
describing interaction of virtual  and real photon with
creation of a  charged fermion pair  for various chiral states
are considered in details. Similar results are presented
for Compton subprocess with virtual photon.

 Contribution of emission of virtual, soft,
and hard real additional photons are taken into account
explicitly.
The relevant cross sections  expressed in terms of impact factors
are in agreement with structure function approach
in the leading logarithmical approximation.
 Contributions of the next to leading
terms are presented in an analytical form.
Accuracy estimation is discussed.
\end{abstract}

\maketitle

\section{Introduction}

\begin{figure}
\includegraphics[scale=0.8]{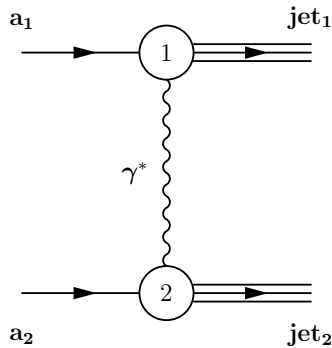}
\caption{ General diagram for the process.}
\label{genfig}
\end{figure}

\begin{figure}
\includegraphics[scale=0.8]{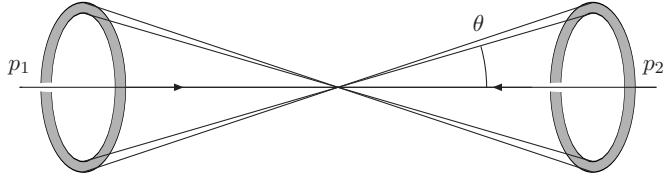}
\caption{ Kinematics of quasiperipheral process.}
\label{fig1}
\end{figure}

Much  attention was paid during the  last decades (see \cite{serbo} and references therein) to
different processes of kind
\ba
 a_1(p_1,\delta_1)+a_2(p_2,\delta_2)\to jet_1^{(\lambda_1)}+jet_2^{(\lambda_2)}, \qquad a_{1,2}=e^\pm,\gamma; \qquad
(p_1+p_2)^2=s\gg m_i^2,
\ea
and $\delta_i,(\lambda_i)$ describe the polarization states of the initial and jet particles.
Below we choose $\delta_1=\delta_2=+1$ without loss of generality (see fig \ref{genfig}).
These processes can be studied at high energy collisions of the initial particles in
peripheral kinematics, i.e. small angles $\theta$ of emission of
jet particles to the direction of their parent particle (center of
mass (cms) frame of initial particles is implied), see Fig. \ref{fig1}.
A remarkable property of nondecreasing  of differential and total cross sections
 on cms total energy $\sqrt{s}$
in this kinematics is commonly known \cite{Baier78}. This property is the
consequence of the presence of massless vector particle (photon) in
the scattering channel state. The contributions of Feynman
diagrams with fermions as well as the interference of amplitudes of
these kinds is suppressed compared with photon exchange ones.

As the corresponding cross sections of the relevant QED
processes are large numerically, they provide an essential
background when studying the effects of weak and strong
interactions. Besides, these processes can be used for monitoring
and calibration purposes.

Unfortunately, very small emission angles can not be measured when using in practice.
So we suggest  considering  the
processes (1) in the so-called quasiperipheral (QP) kinematics
which implies the values of emission angles to be small compared with
unity  but much bigger than $m/E=2m/\sqrt{s}$, with $m$
is characteristic mass of jet's constituents:
\begin{equation}
\label{accur}
\frac{2m_i}{\sqrt{s}}\ll\theta_i \ll 1,\qquad m_i^2\ll-q^2\ll s,\quad q=-p_1+\sum
p_{i1}=-\sum p_{i2}+p_2,
 \end{equation}
with $p_{i1},p_{i2}$-4-momenta of particles from $jet_{1,2}$ and $q$
the momentum  of the t-channel virtual photon. The QP kinematics
provides the independence on energy of differential cross sections
but has accuracy of an order of $\theta^2$- the order of
contributions of neglected terms compared to ones considered.

Another important property of QP kinematics is the independence of spin
states of $a_1-jet_1$ and $a_2-jet_2$ blocks of a process. This is a reason why we can put
$\delta_{1,2}=+1$. This fact can be seen by using
Gribov's form of Green function of exchanged
photon with momentum $q$
\ba
\frac{g_{\mu\nu}}{q^2}=\frac{1}{q^2}[g_{\mu\nu\bot}+\frac{2}{s}[p_{1\mu}p_{2\nu}+p_{2\mu}p_{1\nu}]],
\ea
which results in form of the amplitude

\begin{equation}
\label{M12}
M^{(12)}=-i\frac{4\pi\alpha}{q^2} g_{\mu\nu}(q)J^{1\mu}(q)J^{2\nu}(-q)=
\frac{8\pi\alpha i s}{-q^2}\Phi_1\Phi_2,
\end{equation}
with $J^{1,2}$-currents associated with  blocks 1 and 2 of Feynman
diagram, Fig. \ref{genfig}, and their light-cone projections (LP) defined as
\begin{equation}
\Phi_1=\frac{1}{s}J^{1,\mu}p_{2\mu},\qquad \Phi_2=\frac{1}{s}J^{2,\nu}p_{1\nu}.
\end{equation}
The LP factors $\Phi^i$ do not depend on $s$ in the limit $s\to \infty$.

At this stage, we introduce Sudakov's parameterization of 4-momenta
\begin{gather}
p_{i1}=\alpha_i p_2+x_i p _1+p_{\bot i 1},\qquad\sum x_i=1,
\qquad\alpha_i=\frac{\vec{p}_{i 1}^{\,\,\,2}}{sx_i}, \nn
p_{j2}=y_jp_2+\beta_jp_1+p_{\bot j 2},\qquad \sum y_j=1,
\qquad \beta_j=\frac{\vec{p}^{\,\,\,2}_{j 2}}{sy_j}, \nn q=\alpha p_2+\beta
p_1+q_\bot, \qquad q^2\approx-\vec{q}^{\,\,\,2},\qquad p_{\bot j}^2=-\vec{p}_j^{\,\,\,2}, \nn
\sum \vec{p}_{i 1}=\vec{q},\qquad \sum \vec{p}_{i 2}=-\vec{q}.
\end{gather}

Sudakov's longitudinal parameters $\alpha,\beta$ of the exchanged photon
with momentum $q$ are related with jet's invariant mass squares
\be
s_1=(q+p_1)^2\approx -\vec{q}^{\,\,\,2}+s\alpha,\qquad s_2=(-q+p_2)^2\approx -\vec{q}^{\,\,\,2}-s\beta.
\ee

Here we use the on-mass shell condition
$p_{i1}^2=p_{j2}^2=0$, conservation law, and introduce the
Euclidean 2-dimensional vectors $(p_{i\bot} p_{1,2})=0$.

We should like to note that the current conservation condition $J^{1\mu}q_\mu= J^{2\nu}q_\nu=0$
leads to
\be
\frac{1}{s}p_2^\mu J_\mu^{1} =- \frac{1}{s\alpha} q_\bot^\mu  {J}^{1\mu},\quad
\frac{1}{s}p_1^\nu J_\nu^{2} =- \frac{1}{s\beta} q_\bot^\mu  {J}^{2\mu}.
\ee
 This property of turning to zero of matrix elements at small $\vec{q}$  we use as an important
check of calculations (see (\ref{M12})).

The differential cross section can be written in terms of the Cheng-Wu impact factors
\cite{chengwu}:
\begin{gather}
\dd\sigma^{(12)}=\frac{\alpha^2}{\pi^2}\frac{\dd^2\vec{q}}{(\vec{q}^{\,\,\,2})^2}\int \dd\tau_1^{(\lambda_1)}
\int \dd\tau_2^{(\lambda_2)}, \nonumber \\
\int\dd\tau_i^{(\lambda_i)}=\int|\Phi_i^{(\lambda_i)}(q)|^2\dd\Gamma_{i}, \quad i=1,2\,,
\end{gather}
with
\begin{gather}
\dd\Gamma_1=(2\pi)^4\int \dd s_1\delta(p_1+q-\sum p_{i1})\prod
\frac{\dd^3p_{i1}}{2\varepsilon_{i1}(2\pi)^3}, \nonumber \\
\dd\Gamma_2=(2\pi)^4\int \dd s_2\delta(p_2-q-\sum p_{i2})\prod
\frac{\dd^3p_{i2}}{2\varepsilon_{i2}(2\pi)^3}.
\end{gather}

Impact factors $\int\dd\tau_i$ do not depend on $s$.
For the cases when a jet consist
of one,  two or three  particles we have
\begin{gather}
\int \dd\Gamma_1^{(1)}=2\pi, \nn
\dd\Gamma_1^{(2)}=\frac{\dd^2\vec{p}_{11}\dd x_1}{2(2\pi)^2x_1x_2}, \qquad
\vec{p}_{11}+\vec{p}_{21}=\vec{q},\qquad x_1+x_2=1,
\nn
\dd\Gamma_1^{(3)}=\frac{\dd^2\vec{p}_{11}\dd^2 \vec{p}_{21}\dd x_1\dd x_2}{4(2\pi)^5 x_1x_2x_3}, \qquad x_1+x_2+x_3=1,
\qquad \vec{p}_{11}+\vec{p}_{21}+\vec{p}_{31}=\vec{q}.
\end{gather}

For conversion of
the initial photon with momentum $p_1$ and chirality $\lambda$ to the charged
fermion-antifermion pair
\ba
\label{phot_proc}
\gamma^*(q)+\gamma(p_1,\lambda)\to e^+(q_+)+e^-(q_-,\sigma)
\ea
we accept the description of chiral states of photon and lepton, developed in \cite{calcul}:
\begin{gather}
\hat{e}^{\lambda}(p_1)=N_\gamma(\hat{q}_-\hat{q}_+\hat{p}_1\omega_{-\lambda}-\hat{p}_1\hat{q}_-\hat{q}_+\omega_\lambda),
\qquad (e^\lambda)^2=0,\qquad (e^\lambda e^{-\lambda})=-1,
\nn
N^2_\gamma=\frac{2}{s_1\chi_+\chi_-},\quad s_1=2q_+q_-,\quad \chi_{\pm}=2p_1q_{\pm},
\quad \omega_\sigma=\frac{1+\sigma\gamma_5}{2},\quad \sigma=\pm 1.
\label{phot_desig}
\end{gather}

Chiral states of fermions are defined as $
u^\sigma=\omega_{-\sigma} u,v^\sigma=\omega_\sigma v$.
Hereafter we will imply that chiral states of the  subprocess (\ref{phot_proc})
are defined as amplitudes with a definite chiral state $\Phi^{\lambda\sigma}$ of the initial photon ($\lambda$) and
one electron  ($\sigma$) from pair.
The LP factor of the photon $\Phi^{\lambda\sigma}$ in the Born
approximation has the form:
\begin{gather}
\Phi^{\gamma,++}_B= N_\gamma f_0\bar{u}(q_-)\omega_-\hat{q}_+\hat{q}\frac{\hat{p}_2}{s}\omega_+v(q_+),\\ \nn
\Phi^{\gamma,+-}_B= N_\gamma f_0\bar{u}(q_-)\frac{\hat{p}_2}{s}\hat{q}\hat{q}_-\omega_-v(q_+), \qquad f_0=-i\sqrt{4\pi\alpha}.
\end{gather}
Note that in the combination $\hat{p}_2\hat{q}=\hat{p}_2\hat{q}_\bot$ we can consider
4-vector $q$ as a two-dimensional one $q_\mu=q_{\bot\mu}$.

The property of the LP factors $\Phi_{1,2}(\vec{q})\to 0$ at $|\vec{q}|\to 0$ is the consequence of gauge invariance,
 as we have note above.

The relevant impact factors are
\ba
\int \dd\tau_B^{\gamma,+\pm}=\frac{\alpha}{\pi}\int\frac{d^2\vec{q}_- dx_-}{x_+x_-}\frac{\vec{q}^{\,\,\,2}x^2_\pm}{\chi_+\chi_-}, \nonumber \\
\chi_\pm=\frac{\vec{q}^{\,\,\,2}_\pm}{x_\pm},\quad x_++x_-=1,\quad \vec{q}_++\vec{q}_-=\vec{q}\;.
\ea

The LP amplitudes $\Phi_B^{\gamma,-\pm}$ and the corresponding impact factors can be obtained by application of
the space reflection operator.

Let us consider  the pair production process by photon on electron in the case of
definite chiral states of all the particles
\ba
\gamma(p_1,\lambda=+)+e^-(p_2,\eta) \to e^+(q_+,\mp)+e^-(q_-,\pm)+ e^-(p_2',\eta).
\ea

Using the impact factor of electron-spectator in the lowest order of
perturbation theory (PT) with a single-particle jet
$e^-(p_2,\eta)+\gamma^*(-q)\to e^-(p_2')$:
\ba
\Phi_2^\eta=\bar{u}(p_2')\frac{\hat{p}_1}{s}\omega_\eta
u(p_2),\qquad |\Phi^\eta_2|^2=1,\qquad \int d\tau_2=2\pi,
\ea
we obtain the cross section of pair
photo-production on electron with definite chiral states of the initial
photon and positron from the pair (it does not depend on the chiral
state of the spectator)
\begin{gather}
\frac{\dd\sigma^{\gamma,++}_{B,\eta}}{\dd\Gamma}=
\frac{\dd\sigma^{\gamma,--}_{B,\eta}}{\dd\Gamma}=\frac{2x_+^2\alpha^3}{\pi^2\vec{q}^{\,\,\,2}\chi_+\chi_-}\,,
\qquad
\frac{\dd\sigma^{\gamma,+-}_{B,\eta}}{{\dd\Gamma}}=
\frac{\dd\sigma^{\gamma,-+}_{B,\eta}}{\dd\Gamma}=\frac{2x_-^2\alpha^3}{\pi^2\vec{q}^{\,\,\,2}\chi_+\chi_-}\,,
\nn
\dd\Gamma=\frac{\dd^2\vec{q}\,\dd^2\vec{q}_-\dd x_-}{x_+x_-}\;.
\end{gather}

Also, we put here for consideration the process of single photon emission  at high energy electrons peripherical scattering
\ba
e^-(p_2,\eta)+e^-(p_1,\sigma=+)\to e^-(p_1',+)+\gamma(k_1,\lambda=\pm)+e^-(p_2',\eta),
\ea
with definite chiral states
\begin{gather}
\hat{e}^{\lambda}(k_1)=N_1(\hat{p}_1'\hat{p}_1\hat{k}_1\omega_{-\lambda}-\hat{k}_1\hat{p}_1'\hat{p}_1\omega_{\lambda}),
\nn
N^2_1=\frac{2}{u\chi\chi'},\qquad u=2p_1p_1',\qquad \chi=2p_1k_1,\qquad \chi'=2p_1'k_1.
\end{gather}
The LP factor of the electron $\Phi^{e,\sigma\lambda}$ is
\ba
\Phi^{e,++}_B=N_1 f_0\bar{u}(p_1')\omega_-\hat{p}_1\hat{q}\frac{\hat{p}_2}{s}\omega_+u(p_1)\,,
\qquad
\Phi^{e,+-}_B=-N_1 f_0\bar{u}(p_1')\frac{\hat{p}_2}{s}\hat{q}\hat{p}_1'\omega_+u(p_1)\,.
\ea
The corresponding impact factors are
\ba
\int\dd\tau_B^{e,++}=\frac{\alpha}{\pi}\int\frac{\dd^2\vec{k}_1\, \dd x_1}{x_1x'}\frac{\vec{q}^{\,\,\,2}}{\chi\chi'},
\qquad
\int\dd\tau_B^{e,+-}=\frac{\alpha}{\pi}\int\frac{\dd^2\vec{k}_1\, \dd x_1}{x_1x'}\frac{\vec{q}^{\,\,\,2}(x')^2}{\chi\chi'}\;,
\ea
where $x_1,x'$-are the energy fractions of photon and the scattered electron from the jet.
From the conservation law and on-mass shell conditions we have
\ba
x_1+x'=1,\quad
\vec{p}_1\,'+\vec{k}_1=\vec{q},\quad
\chi=\frac{\vec{k}_1^2}{x_1},\quad
\chi'=\frac{1}{x_1x'}(x_1\vec{p}-x'\vec{k}_1)^2.
\ea

The differential cross sections have the form:
\begin{gather}
\frac{\dd\sigma^{e,++}_{B,\eta}}{\dd\Gamma}=
\frac{\dd\sigma^{e,--}_{B,\eta}}{\dd\Gamma}=
\frac{2\alpha^3}{\vec{q}^{\,\,\,2}\pi^2\chi\chi'}\;,\qquad
\frac{\dd\sigma^{e,+-}_{B,\eta}}{\dd\Gamma}=
\frac{\dd\sigma^{e,-+}_{B,\eta}}{\dd\Gamma}=\frac{2\alpha^3(x')^2}{\vec{q}^{\,\,\,2}\pi^2\chi\chi'}\;,\nn
\dd\Gamma=\frac{\dd^2 \vec{k}_1\,\dd x_1\dd^2 \vec{q} }{x_1x'}\;.
\end{gather}

Our paper is
arranged in the following way. In Sections 2 we consider the virtual
(in the one-loop approximation) and soft real photon emission
contribution to the photon impact factor. In Section 3
the similar calculations are presented for the electron impact factor. In Sections 4 and 5
we consider the emission of
an additional hard photon in collinear and non-collinear kinematics.
Some general remarks are given in Conclusion.
We discuss in particular the validity of the structure
function approach in the leading and next-to-leading approximations.
 The relevant one-loop integrals are listed in Appendix A. Appendix B contains the explicit expressions for
non-leading contributions arising from virtual and soft real photon emission.
These nonleading contributions expressed in terms of K-factor turn out to be
the quantities of an order of unity for typical experimental conditions.

\section{photon impact factor: virtual and soft photon contribution}

\begin{figure}
\includegraphics[scale=0.8]{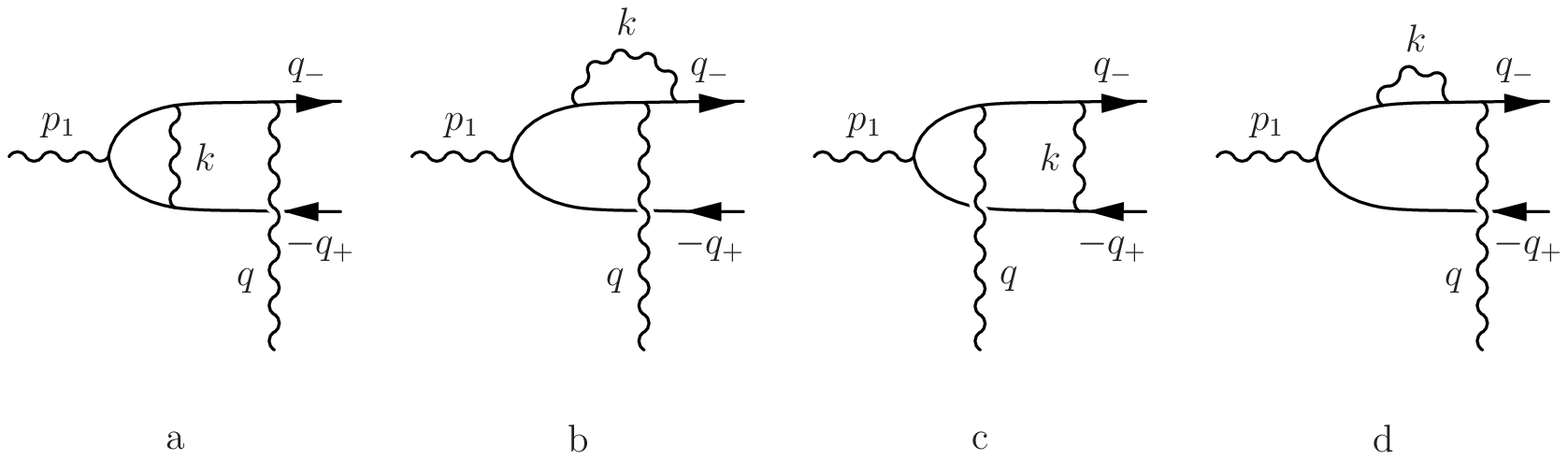}
\caption{ Photon impact factor diagram (I)}
\label{figI}
\end{figure}

\begin{figure}
\includegraphics[scale=0.8]{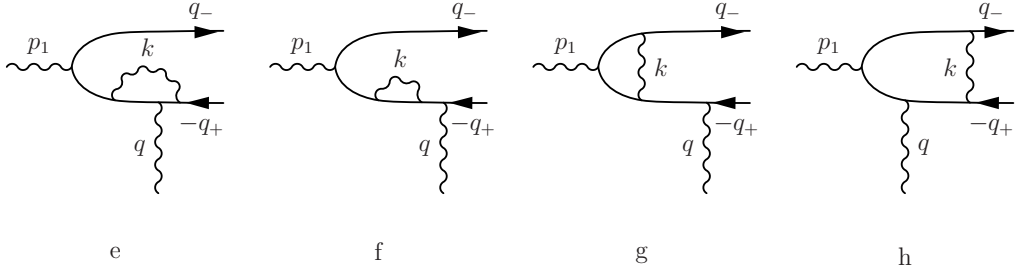}
\caption{ Photon impact factor diagram (II)}
\label{figII}
\end{figure}

We can all diagrams (see fig \ref{figI},\ref{figII})  divide into several types,
some of which (fig \ref{figI} a,d, and \ref{figII} e,h)  can be obtained by simple exchanges of
chirality and 4-momenta of particles:
\ba
\label{change}
Re[\Phi^{\gamma,+\pm}_{+,\,i}(\Phi^{\gamma,+\pm}_{Born})^*]=Re[\Phi^{\gamma,+\mp}_{-,\,i}(\Phi^{\gamma,+\mp}_{Born})^*(q_-\to q_+, q_+\to q_-)],
\ea
with $i=\Sigma V,V, B $-the self energy, vertex, and box-type Feynman diagram contribution.

Here the low subscript describes the absorbtion of virtual photon by
electron (-) (Fig. \ref{figI} a-d) or positron (+) line (Fig. \ref{figII} e-h).

One class of RC to electron impact factor consists of  the renormalized electron
mass operator and the vertex function with only one off-mass shell electron or positron
(see Fig. \ref{figI} a,d and \ref{figII} f,g). Its contribution can be written in the form \cite{ahber}
\ba
\Phi^{\gamma,\lambda\sigma}_{-, \Sigma V}&=&\frac{(4\pi\alpha)^{3/2}}{16\pi^2}\bar{u}(q_-)\frac{\hat{p}_2}{s}
\frac{\hat{p}_1-\hat{q}_+}{-\chi_+}
 \nonumber \\
&\times&\biggl[
2(\frac{3}{2}-l_l+\frac{1}{2}l_+)\hat{e}^\lambda
+\int\frac{\dd^4k}{i\pi^2}\frac{\gamma^\mu(-\hat{q}_++\hat{p}_1-\hat{k})
\hat{e}^\lambda(-\hat{q}_+-\hat{k})\gamma^\mu}{(0)(\bar{2})(q)}
\biggr]
\omega^\sigma v(q_+),
\nonumber \\
\Phi^{\gamma,\lambda\sigma}_{+, \Sigma V}&=&\frac{(4\pi\alpha)^{3/2}}{16\pi^2}\bar{u}(q_-)
\biggl[
2(\frac{3}{2}-l_l+\frac{1}{2}l_-)\hat{e}^\lambda
+\int\frac{\dd^4k}{i\pi^2}
\frac{\gamma^\mu(\hat{q}_--\hat{k})\hat{e}^\lambda(\hat{q}_--\hat{p}_1-\hat{k})\gamma^\mu}{(0)(2)(\bar{q})}
\biggr]
 \nonumber \\
&\times&\frac{\hat{q}_--\hat{p}_1}{-\chi_-}\frac{\hat{p}_2}{s}
\omega^\sigma v(q_+),
\qquad l_l=\ln\frac{m^2}{\lambda^2}, \quad l_\pm=\ln\frac{\chi_\pm}{m^2},
\label{vertel}
\ea
with the denominators $(0),(2),(\bar{2}),(q),(\bar{q})$ are defined below (see Appendix A).
After integration we obtain:
\ba
\Phi^{\gamma,+-}_{-, \Sigma V}=-f_0N_\gamma\frac{\alpha}{2\pi}(l_+-\frac{1}{2})
\bar{u}(q_-)\frac{\hat{p}_2}{s}\hat{p}_1\hat{q}_-\omega_+v(q_+),
\nonumber \\
\Phi^{\gamma,++}_{+, \Sigma V}=f_0N_\gamma\frac{\alpha}{2\pi}(l_--\frac{1}{2})
\bar{u}(q_-)\hat{q}_+\hat{p}_1\frac{\hat{p}_2}{s}\omega_-v(q_+).
\ea

After multiplying the relevant Born amplitude  we obtain \cite{KF85}:
\ba
2\Phi^{\gamma,+\pm}_{\mp, \Sigma V}(\Phi^{\gamma,+\pm}_{Born})^*=
\frac{8\alpha^2}{\chi_+\chi_-}x_\pm(\vec{q}\vec{q}_\pm)(l_\mp-\frac{1}{2}).
\ea

We should like to note that the $\Sigma V$ contribution does not satisfy the gauge
condition (turning to zero as $\vec{q}^{\,\,\,2}$ at small $\vec{q}$).
We will see later that the total sum satisfies the  gauge condition.

The contribution of the vertex functions with a virtual photon can be written  in the form:
\ba
\Phi^{\gamma,\lambda\sigma}_{-,V}=\frac{(4\pi\alpha)^{3/2}}{16\pi^2}
\int\frac{\dd^4k}{i\pi^2}\frac{
\bar{u}(q_-)\gamma_\mu(\hat{q}_--\hat{k})\frac{\hat{p}_2}{s}(\hat{q}_--\hat{q}-\hat{k})
\gamma_\mu(\hat{p}_1-\hat{q}_+)\hat{e}^\lambda\omega^\sigma v(q_+)}{(0)(2)(q)(-\chi_+)}\;,
\nonumber \\
\Phi^{\gamma,\lambda\sigma}_{+,V}=\frac{(4\pi\alpha)^{3/2}}{16\pi^2}
\int\frac{\dd^4k}{i\pi^2}\frac{
\bar{u}(q_-)\hat{e}^\lambda(\hat{q}_--\hat{p}_1)\gamma_\mu
(-\hat{q}_++\hat{q}-\hat{k})\frac{\hat{p}_2}{s}(-\hat{q}_+-\hat{k})
\gamma_\mu\omega^\sigma v(q_+)}{(0)(\bar{2})(\bar{q})(-\chi_-)}\;.
\label{virthardel}
\ea
Using the list of integrals (see Appendix for designations) we obtain
\begin{gather}
2\Phi^{+-}_{-,V}(\Phi^{+-}_B)^*=-2|\Phi_B^{+-}|^2 \frac{\alpha}{2\pi}
\biggl[-\frac{1}{2}L-\frac{1}{4}-\frac{\chi_++2\vec{q}^{\,\,\,2}}{2a}l_++
\frac{3\vec{q}^{\,\,\,2}}{2a}l_q-\vec{q}^{\,\,\,2}J_{02q}\biggr],
\nonumber \\
2\Phi^{++}_{+,V}(\Phi^{++}_B)^*=-2|\Phi_B^{++}|^2 \frac{\alpha}{2\pi}
\biggl[-\frac{1}{2}L-\frac{1}{4}-\frac{\chi_-+2\vec{q}^{\,\,\,2}}{2\tilde{a}}l_-+
\frac{3\vec{q}^{\,\,\,2}}{2\tilde{a}}l_q-\vec{q}^{\,\,\,2}J_{0\bar{2}\bar{q}}\biggr],
\nonumber  \\
a=\chi_+-\vec{q}^{\,\,\,2}, \quad  \tilde{a}=\chi_--\vec{q}^{\,\,\,2}
\end{gather}
The other contributions $\Phi^{++}_{-,V}(\Phi^{++}_B)^*$, $\Phi^{+-}_{+,V}(\Phi^{+-}_B)^*$,
$\Phi^{\gamma,+-}_{+, \Sigma V}(\Phi^{+-}_B)^*$,$\Phi^{\gamma,++}_{-, \Sigma V}(\Phi^{++}_B)^*$
is equal to zero.
We remind that we work in the framework of the unrenormalized field theory. The regularization procedure
consists in the replacement of ultra-violet cut-off logarithm $L=\ln\frac{\Lambda^2}{m^2}$ as
 $L\to 2l_l-\frac{9}{2}$ (see \cite{ahber}).

The most complicated case is the calculation of the box-type contribution.
It can be written down in the form (see Fig. \ref{figI} c, Fig. \ref{figII} h)
\ba
\Phi^{\gamma,\lambda\sigma}_{-, box}=
\frac{(4\pi\alpha)^{3/2}}{16\pi^2}\int\frac{\dd^4k}{i\pi^2}\frac{\bar{u}(q_-)\gamma^\mu
(\hat{q}_--\hat{k})\frac{\hat{p}_2}{s}
(\hat{q}_--\hat{q}-\hat{k})\hat{e}^\lambda(-\hat{q}_+-\hat{k})
\gamma^\mu\omega^\sigma v(q_+)}{(0)(\bar{2})(2)(q)},
\nonumber \\
\Phi^{\gamma,\lambda\sigma}_{+, box}=
\frac{(4\pi\alpha)^{3/2}}{16\pi^2}\int\frac{\dd^4k}{i\pi^2}\frac{\bar{u}(q_-)
\gamma^\mu(\hat{q}_--\hat{k})\hat{e}^\lambda
(\hat{q}_--\hat{p}_1-\hat{k})\frac{\hat{p}_2}{s}
(-\hat{q}_+-\hat{k})
\gamma^\mu\omega^\sigma v(q_+)}{(0)(\bar{2})(2)(\bar{q})}.
\label{boxel}
\ea
All the details about loop calculations and relevant integrals  can be found
in Appendix A.
It is worth mentioning that in the case of the box-type contribution
both chiral amplitudes $\lambda=+1$, $\sigma=\pm1$ are nonzero.

An additional real soft photon emission contribution to the LP factor has the standard form
\ba
\Phi^{\gamma,\lambda\sigma\eta}_{soft}=\Phi^{\gamma,\lambda\sigma}_B\sqrt{4\pi\alpha}(\frac{q_-}{q_-k}-\frac{q_+}{q_+k})e(k)^{\eta}.
\ea
The corresponding contribution to the impact factor is:
\ba
\int\frac{\dd^3k}{16\omega\pi^3}\sum\limits_\eta|\Phi^{\gamma,\lambda\sigma\eta}_{soft}|^2|_{\omega<\Delta \varepsilon\ll \varepsilon_\gamma}.
\label{soft_gamma}
\ea
Here $\varepsilon_\gamma$ is the energy of the initial electron in c.m.s.
The result is
\ba
\dd\tau_{ soft}^{\gamma,\lambda\sigma}=
\frac{\alpha}{\pi}\dd\tau_B^{\gamma,\lambda\sigma}\biggl[(l_s-1)(l_l+\ln\frac{(\Delta)^2}{x_+x_-})+\frac{1}{2}l_s^2
-\frac{1}{2}\ln^2\frac{x_+}{x_-}-\frac{\pi^2}{6}\biggr],
 \\ \nonumber
 \Delta=\frac{\Delta \varepsilon}{\varepsilon_\gamma}, \quad l_s=\ln\frac{s_1}{m^2}.
\ea
We use here the smallness of the angle between 3-momenta of pair components in beams the center-of-mass frame.
Smallness of emission angles provides the possibility to perform angular integration in (\ref{soft_gamma})
in frame $S_0$ which coinside with cms frame \cite{MaxT}.

After summing all contributions (\ref{vertel},\ref{virthardel},\ref{boxel})
and adding the soft photon contribution we explicitly see
the cancellation of an auxiliary parameter $\lambda$ and squared large logarithm:
\ba
2[\dd\tau^{\gamma,+\pm}_{\pm, box}+\dd\tau^{\gamma,+\pm}_{\mp,\Sigma V}+\dd\tau^{\gamma,+\pm}_{\mp,V}]
+\dd\tau_{soft}^{\gamma,+\pm}
 \nonumber \\
=\frac{\alpha}{2\pi}\dd\tau_B^{\gamma\pm}[(l_s-1)(4\ln\Delta+3-2\ln(x_+x_-))+K^{\gamma,\pm}_{SV}].
\label{taugam}
\ea

Here we can see that leading logarithm contribution (containing the factor $(l_s-1)$) is proportional
to the Born cross section, so our calculation is in agreement with
the structure function approach predictions, namely the leading logarithm contribution
is exactly the $\Delta$ part of the evolution equation kernel (see (\ref{kernelP})).
All nonleading terms are gathered in the so-called
K-factor.

Due ro gauge-invariance the right part of (\ref{taugam}) including
K-factor turns to zero as $\vec{q}^{\,\,\,2}\to0$. This fact provides an important check
of our calculation.

The $K^{\gamma,\pm}_{SV}$ is presented in the analytical form in Appendix B.

The contribution from emission of hard photon which eliminate the $\Delta$-dependence
can be written as a sum of two parts, one from collinear and
the other from noncollinear kinematics. It will be considered below.

\section{electron impact factor: virtual and soft photon contribution}

\begin{figure}
\includegraphics[scale=0.8]{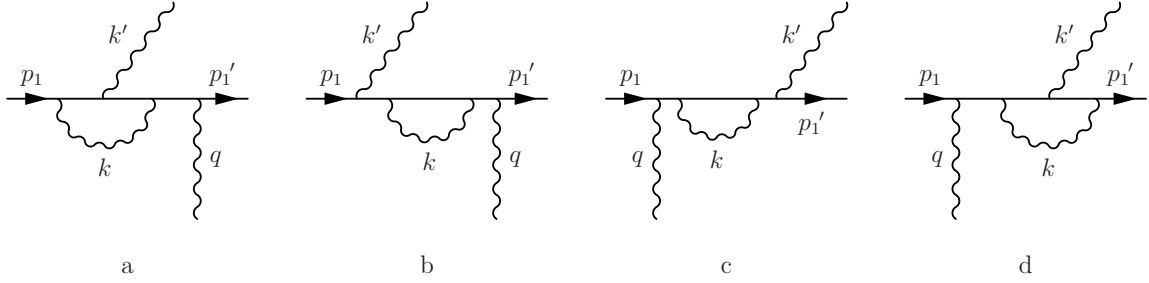}
\caption{Electron impact factor diagrams (I)}
\label{figeI}
\end{figure}

\begin{figure}
\includegraphics[scale=0.8]{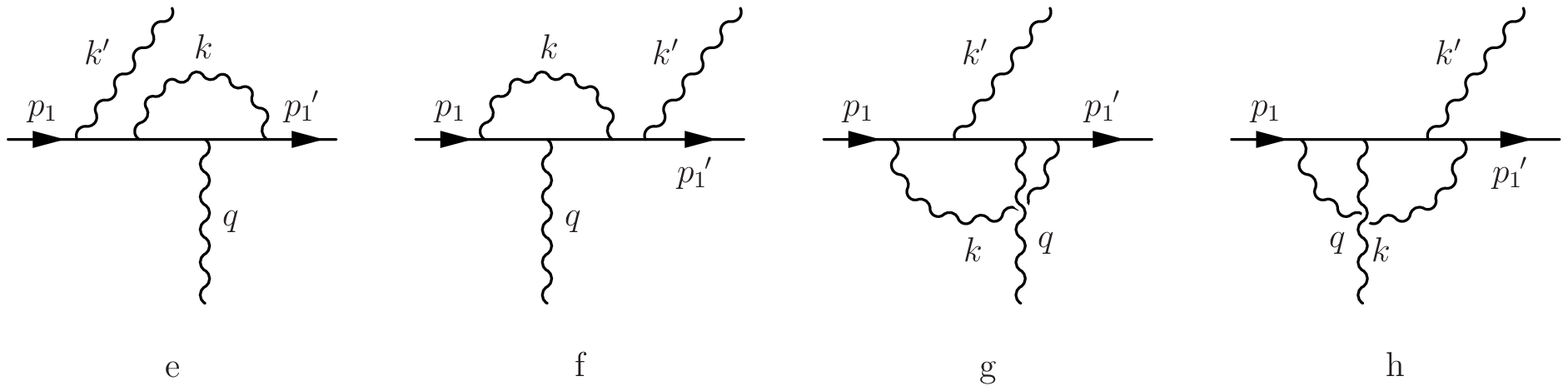}
\caption{Electron impact factor diagrams (II)}
\label{figeII}
\end{figure}

In the same way we calculate the electron impact factor.
All diagrams (see Fig \ref{figeI},\ref{figeII}) are divided into six types, the contribution of three of them to LP
(Fig \ref{figeI} c,d and \ref{figeII} f,h)
can be obtained by a simple exchange
\ba
\label{change_electron}
Re[\Phi^{e,+\mp}_{i,contr}(\Phi^{e,+\mp}_{Born})^*]=Re[\Phi^{e,+\pm}_{f, contr}(\Phi^{e,+\pm}_{Born})^*(p_1\to -p_1', p_1'\to -p_1)],
\ea
where the low subscript correspond to interaction of virtual
photon with the initial (i) or scattered (f) electron, and
$contr=\Sigma V,V, box $-the self energy, vertex, and box-type Feynman diagram contributions.

We take into account the contributions from
self-energy (Fig. \ref{figeI} a,b), vertex (Fig. \ref{figeII}e) and box-type (fig. \ref{figeII}g) FD amplitudes
\ba
\label{elect_all_contr}
\Phi^{e,+\lambda}_{i, \Sigma V}&=&\frac{(4\pi\alpha)^{3/2}}{16\pi^2(-\chi)}\bar{u}(p_1')
\frac{\hat{p}_2}{s}(\hat{p}_1-\hat{k}_1)
\\ \nonumber
&\times&\biggl[
\int\frac{\dd^4k}{i\pi^2}\frac{\gamma_\mu(\hat{p}_1-\hat{k}_1-\hat{k})
\hat{e}^{\lambda}(\hat{p}_1-\hat{k})\gamma_\mu}{(0)_e(1)_e(q)_e}
 +2(\frac{1}{2}\ln\frac{\chi}{m^2}-2\ln\frac{m}{\lambda}+\frac{3}{2})\hat{e}^{\lambda}\biggr]
\omega_+u(p_1),
\\ \nonumber
\Phi^{e,+\lambda}_{i, V}&=&\frac{(4\pi\alpha)^{3/2}}{16 \pi^2(-\chi)}
\int\frac{\dd^4k}{i\pi^2}\frac{\bar{u}(p_1')\gamma_\mu(\hat{p}_1'-\hat{k})\frac{\hat{p}_2}{s}
(\hat{p}_1'-\hat{k}-\hat{q})\gamma_\mu
(\hat{p}_1-\hat{k}_1)e^\lambda\omega_+ u(p_1)}{(0)_e(2)_e(q)_e},
\\ \nonumber
\Phi^{e,+\lambda}_{i, box}&=&\frac{(4\pi\alpha)^{3/2}}{16\pi^2}
\int\frac{\dd^4k}{i\pi^2}\frac{\bar{u}(p_1')\gamma_\mu(\hat{p}_1'-\hat{k})\frac{\hat{p}_2}{s}
(\hat{p}_1'-\hat{k}-\hat{q})\hat{e}^\lambda
\gamma_\mu\omega_+ u(p_1)}{(0)_e(1)_e(2)_e(q)_e}.
\ea

The first two contribute (see details in Appendix):
\ba
2\Phi^{e,+-}_{i,V}(\Phi_B^{e,+-})^*&=&-2|\Phi_B^{e,+-}|^2\frac{\alpha}{2\pi}
\biggl[-\frac{1}{2}L-\frac{1}{4}-\vec{q}^{\,\,\,2}I_{(0)_e(2)_e(q)_e}
+\frac{3\vec{q}^{\,\,\,2}}{2d}\ln\frac{\vec{q}^{\,\,\,2}}{m^2}-\frac{\chi+2\vec{q}^{\,\,\,2}}{2d}\ln\frac{\chi}{m^2}\biggr],
 \nonumber \\
2\Phi^{e,+-}_{i,\Sigma V}(\Phi_B^{e,+-})^*&=&\frac{8\alpha^2}{\chi\chi'}
(\ln\frac{\chi}{m^2}-\frac{1}{2})x'[x'\vec{k}_1-x\vec{p}_1\,']\vec{q},
 \quad
d=\chi-\vec{q}^{\,\,\,2},\quad
l_{\chi'}=\ln\frac{\chi'}{m^2}-i\pi.
\label{electr_part_contr}
\ea
The contribution for other polarization can be obtained by
substitution (\ref{change_electron}).

The soft photon contribution has the  standard form
(the energy of soft photon does not exceed $\Delta \varepsilon$)
\ba
\dd\tau_{soft}^{e,+\pm}=\dd\tau_B^{e,+\pm}\frac{\alpha}{\pi}
[(\ln\frac{u}{m^2}-1)(2\ln\frac{m}{\lambda}+2\ln\Delta-\ln x')
+\frac{1}{2}\ln^2\frac{u}{m^2}-\frac{1}{2}\ln^2x'-\frac{\pi^2}{6}],
\ea
where $\Delta=\frac{\Delta\varepsilon}{\varepsilon}$, $\varepsilon$ is the energy of the initial electrons
in cms frame.
We can express the contribution to electron impact  factors with
a definite chiral state:
\ba
2(\dd\tau^{e,+\pm}_{i,\Sigma V}+\dd\tau^{e,+\pm}_{i,V}
+\dd\tau^{e,+\pm}_{i,box}+\dd\tau^{e,+\pm}_{f,box})+\dd\tau_{e,soft}^{+\pm}
\\ \nonumber
=\dd\tau_B^{e,+\pm}\frac{\alpha}{2\pi}[(\ln\frac{u}{m^2}-1)
(4\ln\Delta+3-2\ln x')+K^{e,+\pm}_{SV}].
\ea
Again here we can see the cancellation of the auxiliary "photon mass" parameter
$\lambda$ and agreement with the prediction of the structure function approach.
The $K^{e,+\pm}_{SV}$ is presented in the  analytical form in Appendix B.

\section{Collinear kinematics of additional hard photon emission contribution}

For appropriate consideration of RC to impact factors, we
have to consider additional hard collinear photon emission. It is convenient to distinguish
the collinear and non-collinear kinematics of emission of hard photon. For this aim we introduce
an auxiliary small parameter $\theta_0\ll 1$.
Collinear kinematics corresponds to the the case when the photon emission
angle $\theta$ to the direction of motion of some charged particle (initial or final) do not exceed $\theta_0$.
Noncollinear kinematics corresponds to large emission angles $\theta>\theta_0$.
Chiral amplitudes in noncollinear
kinematics can be calculated using the methods developed by the CALCUL collaboration \cite{calcul}. The contribution from
collinear kinematics can be obtained using the quasi-real electron method developed in \cite{bfk73}. The total sum
does not depend on the parameter $\theta_0$.
Cancellation of the $\theta_0$ dependence is some check of our calculations.
The nonleading contributions from additional hard photon emission essentially depend on the experimental set-up.
We do include it in the $K$-factors in the structure function picture of impact factors.

Using the quasi-real electron method \cite{bfk73}  for contribution to photon impact factor in collinear kinematics
we obtain:
\ba
d\tau^{\gamma,+\lambda}_{coll}=\frac{\alpha}{2\pi}\int\limits_{x_-(1+\Delta)}^1\frac{dz_-}{z_-}
\biggl[\frac{1+\tilde{x}_-^2}{1-\tilde{x}_-}(l_s+r_-+\ln\theta_0^2-1)+1-\tilde{x}_- \biggr]
d\tau^{\gamma,+\lambda}_B(\frac{q_-}{z_-},q_+) \nonumber \\
+\frac{\alpha}{2\pi}
\int\limits_{x_+(1+\Delta)}^1\frac{dz_+}{z_+}\biggl[\frac{1+\tilde{x}_+^2}{1-\tilde{x}_+}(l_s+r_++\ln\theta_0^2-1)+1-\tilde{x}_+\biggr]
d\tau^{\gamma,+\lambda}_B(q_-,\frac{q_+}{z_+}),
\ea
where the first term in square brackets corresponds to the emission of hard photon along electron
and the second one-along positron from the pair created.
Beside we use the notation $\tilde{x}_\pm=\frac{x_\pm}{z_\pm}$ and
\ba
l_s=\ln \frac{2q_+q_-}{m^2}=\ln\frac{2E^2x_+x_-(1-c_\gamma)}{m^2},\quad r_\pm=\ln\frac{x_\pm}{2x_\mp(1-c_\gamma)},
\ea
the quantity $c_\gamma$ is the cosine of the angle between pair momenta (the center-of-mass frame of colliding beams is implied).
The "shifted" photon impact factor (conservation law reads as $p_1+q=\frac{1}{z_-}q_-+\frac{1}{z_+}q_+$) is
\ba
\dd\tau^{\gamma,+\pm}_B(\frac{q_+}{z_+},\frac{q_-}{z_-},x_\pm)=\frac{\alpha}{\pi}
\frac{\tilde{x}_\pm^2\vec{q}^{\,\,\,2}}{\vec{q}_+^2\vec{q}_-^2}\dd^2q_-\dd\tilde{x}_-,  \quad
\tilde{x}_++\tilde{x}_-=1,\quad \vec{q}=\frac{1}{z_-}\vec{q}_-+\frac{1}{z_+}\vec{q}_+.
\ea

A similar method can be applied to the problem of calculating a contribution from the collinear kinematics of the photon
emission for the impact factor of electron. The result is
\ba
d\tau^{e,+\lambda}_{coll}&=&\frac{\alpha}{2\pi}\int\limits_0^{1-\Delta}dz_1
\biggl[\frac{1+z_1^2}{1-z_1}(l_u+l_1+\ln\theta^2_0-1)+1-z_1\biggr]d\tau^{e,+\lambda}_B(p_1z_1,p_1') \nonumber \\
&+&\frac{\alpha}{2\pi}\int\limits_{x'(1+\Delta)}^1\frac{dz_2}{z_2}\biggl[\frac{1+(\frac{x'}
{z_2})^2}{1-\frac{x'}{z_2}}(l_u+l_2+\ln\theta^2_0-1)
+1-\frac{x'}{z_2}\biggr]
\dd\tau^{e,+\lambda}_B(p_1,\frac{1}{z_2}p_1'),
\ea
where the first term in the square brackets describes the emission from the initial and the second one the emission from the scattered electron.
Here we use the following designations:
\ba
l_u=\ln\frac{2p_1p_1'}{m^2}=\ln\frac{2E^2x'(1-c_e)}{m^2}, \quad
l_1=\ln\frac{z_1^2}{2x'(1-c_e)},\quad l_2=\ln\frac{x'}{2z_2^2(1-c_e)},
\ea
and $c_e$ is the cosine of the angle between the initial and the scattered electrons momenta.

The "shifted" electron impact factor in the Born approximation (the conservation law reads as $z_1p_1+q=\frac{1}{z_2}p_1'+k_1$) is
\ba
d\tau^{e,+\pm}_B(p_1z_1,\frac{1}{z_2}p_1')=\frac{\alpha}{\pi}\frac{\vec{q}^{\,\,\,2}}{\chi\chi'}\eta^\pm\frac{z_2d^2k dx_1}{x_1x'},
\quad  \eta^+=z_1^2, \quad \eta^-=(\frac{x'}{z_2})^2, \nonumber \\
\chi=\frac{z_1}{x}\vec{k}_1^2,\quad
\chi'=\frac{z_2(\vec{p}_1'x_1-\vec{k}_1
\frac{x'}{z_2})^2}{x_1x'}, \quad
x_1+\frac{x'}{z_2}=1,\quad
\vec{q}=\vec{k}_1+\frac{1}{z_2}\vec{p}_1'.
\ea

The terms containing "large " logarithms $l_s-1 ,l_u-1$ will be included in lepton nonsinglet structure functions
in the Drell-Yan form
of impact factors, whereas the rest terms contribute to the relevant $K$-factors.
So  we can rewrite these formulae in the term of structure function approach (chiral indices are suppressed)
for the electron impact factor
\ba
\dd\tau^{e,coll}=\int\limits_0^1\dd z_1\biggl[ P_\theta(z_1)\frac{\alpha}{2\pi}(l_u-1)+\frac{\alpha}{\pi}K^{i,e}_{coll}
\biggr]
\dd\tau^{i,e}(z_1 p_1,p_1')+\dd\tau^{i,e}_{comp}
\nonumber \\
+\int\limits_{0}^1\frac{\dd z_2}{z_2}\biggl[ P_\theta(\frac{x'}{z_2})\frac{\alpha}{2\pi}(l_u-1)
+\frac{\alpha}{\pi}K^{f,e}_{coll}\biggr]\dd\tau^{f,e}(p_1,\frac{p_1'}{z_2})
+\dd\tau^{f,e}_{comp},
\ea
and for the photon impact factor we have
\ba
\dd\tau^{\gamma,coll}=\int\limits_{0}^1\frac{\dd z_-}{z_-}\biggl[ P_\theta(\frac{x_-}{z_-})\frac{\alpha}{2\pi}(l_s-1)
+\frac{\alpha}{\pi}K^{-,\gamma}_{coll}\biggr]\dd\tau(\frac{q_-}{z_-},q_+)+\dd\tau^{-,\gamma}_{comp}
\nonumber \\
+\int\limits_{0}^1\frac{\dd z_+}{z_+}\biggl[ P_\theta(\frac{x_+}{z_+})\frac{\alpha}{2\pi}(l_s-1)
+\frac{\alpha}{\pi}K^{+,\gamma}_{coll}\biggr]\dd\tau(q_-,\frac{q_+}{z_+})+\dd\tau^{+,\gamma}_{comp},
\ea
where
\ba
P_\theta(z)=\frac{1+z^2}{1-z}\theta(1-z-\Delta),
\ea
and nonleading contributions  equal (photon impact factor)
\ba
\dd\tau^{-,\gamma}_{comp}&=&\frac{\alpha}{2\pi}\int\limits_0^1\frac{\dd z_-}{z_-}P_\theta(\frac{x_-}{z_-})\dd\tau_B^\gamma(\frac{q_-}{z_-},q_+)\ln\theta^2_0,
\nonumber \\
K^{-,\gamma}_{coll}&=&\frac{1}{2}\int\limits_{x_-}^1\frac{\dd z_-}{z_-}
P_\theta(\frac{x_-}{z_-})(r_-+1-\tilde{x}_-)\dd\tau_B^\gamma(\frac{q_-}{z_-},q_+),
\nonumber \\
\dd\tau^{+,\gamma}_{comp}&=&\frac{\alpha}{2\pi}\int\limits_0^1\frac{\dd z_+}{z_+}P_\theta(\frac{x_+}{z_+})\dd\tau_B^\gamma(q_-,\frac{q_+}{z_+})\ln\theta^2_0,
\nonumber \\
K^{+,\gamma}_{coll}&=&\frac{1}{2}\int\limits_{x_+}^1\frac{\dd z_+}{z_+}P_\theta(\frac{x_+}{z_+})
(r_+-1-\tilde{x}_+)\dd\tau_B^\gamma(q_-,\frac{q_+}{z_+}),
\label{kcoll_1}
\ea
and (electron impact factor):
\ba
\dd\tau^{i,e}_{comp}&=&\frac{\alpha}{2\pi}\int\limits_0^1\dd z_1P_\theta(z_1)\dd\tau_B^e(z_1p_1,p_1')\ln\theta^2_0,
\nonumber \\
K^{i,e}_{coll}&=&\frac{1}{2}\int\limits_{0}^1\dd z_1P_\theta(z_1)
(l_1+1-z_1)\dd\tau_B^e(z_1p_1,p_1'),
\nonumber \\
\dd\tau^{f,e}_{comp}&=&\frac{\alpha}{2\pi}\int\limits_{x'}^1 \frac{\dd z_2}{z_2}P_\theta(\frac{x'}{z_2})\dd\tau_B^e(p_1,\frac{p_1'}{z_2})\ln\theta^2_0,
\nonumber \\
K^{f,e}_{coll}&=&\frac{1}{2}\int\limits_{x'}^1 \frac{\dd z_2}{z_2}P_\theta(\frac{x'}{z_2})
(l_2+1-\frac{x'}{z_2})\dd\tau_B^e(p_1,\frac{p_1'}{z_2}).
\label{kcoll_2}
\ea
Terms with $\ln\theta_0^2$ will be compensated with  additional noncollinear photon emission.

\section{Noncollinear hard photon emission contribution}

Contribution to the electron impact factor from the channel of the double Compton scattering process
\begin{gather}
e(p_1,\lambda_1)+\gamma^*(q)\to\gamma(k_1,\lambda_1)+\gamma(k_2,\lambda_2)+e(p_1',\lambda_e),
\nonumber \\
u=2p_1p_1', \quad \chi_i=2k_ip_1,\quad \chi_i'=2k_ip_1',
\end{gather}
 with the emission of both the final electrons
outside the narrow cone $\theta>\theta_0$ can be calculated using the chiral amplitude technique \cite{calcul}.
 The result is
\begin{gather}
\dd\tau^{e\gamma\gamma}_{\lambda_e\lambda_1\lambda_2}=
\frac{\alpha^3}{2\pi^2}|m^{\lambda_e}_{\lambda_1\lambda_2}|^2\frac{\dd^2k_1\dd^2k_2\dd x_1\dd x_2}
{x_1x_2x'}|_{\theta_{1,2}>\theta_0}, \nonumber \\
x'=1-x_1-x_2,\quad \vec{q}=\vec{k}_1+\vec{k}_2+\vec{p}_1',
\end{gather}
with
\begin{gather}
|m^+_{++}|^2=\frac{4\vec{q}^{\,\,\,2}u}{\chi_1\chi_2\chi_1'\chi_2'}\quad
|m^+_{--}|^2=\frac{4(x')^2\vec{q}^{\,\,\,2}u}{\chi_1\chi_2\chi_1'\chi_2'}, \quad
|m^+_{-+}|^2=|m^+_{+-}|^2,
\nonumber \\
|m^+_{+-}|^2=\frac{4}{u^2\chi_1\chi_2\chi_1'\chi_2'}Tr\hat{p}_1'B^+_{+-}\omega_+\hat{p}_1\tilde{B}^+_{+-},
\end{gather}
with
\ba
B^+_{+-}=-\frac{1}{(p_1+q)^2}\hat{p}_1\hat{k}_1\hat{p}_1'\hat{p}_1\hat{k}_2(\hat{p}_1+\hat{q})
\frac{\hat{p}_2}{s} -\frac{1}{(p_1'-q)^2}\frac{\hat{p}_2}{s}
(\hat{p}_1'-\hat{q})\hat{k}_1\hat{p}_1'\hat{p}_1\hat{k}_2\hat{p}_1' \nonumber \\
+\hat{p}_1(\hat{p}_1'+\hat{k}_1)\frac{\hat{p}_2}{s}(\hat{p}_1-\hat{k}_2)\hat{p}_1'.
\ea
It was explicitly shown \cite{bbgk} that the quantity $B^+_{+-}$ turns to zero as $|\vec{q}| \to0$,
whose property is the consequence of
gauge invariance implement for the virtual photon with momentum $q$.

For the aim of checking the $\theta_0$ dependence cancellation for the sum of collinear and non-collinear kinematics contributions we put below the
limiting expressions for $|m^+_{ij}|^2$ for the emission of real photon kinematics
\ba
\theta_1>\theta_0,\quad \theta_1\to  \theta_0,\quad \theta_2\gg\theta_0,
\ea
with $\theta_1$ being the angle of emission of photon with momentum $k_1$ to the initial or final electron momentum.
These limiting values are
\ba
(|m^+_{+-}|^2+|m^+_{++}|^2)_{\chi_1\to 0}&=&\frac{4\vec{q}^{\,\,\,2}}{\chi_1}\frac{((x')^2+(1-x_1)^2)}{x_1(1-x_1)^2\chi_2\chi_2'}, \nonumber \\
(|m^+_{+-}|^2+|m^+_{++}|^2)_{\chi_1'\to 0}&=&\frac{4\vec{q}^{\,\,\,2}}{\chi_1'}\frac{x'}{x_1\chi_2\chi_2'}[1+(1-x_2)^2], \nonumber \\
(|m^+_{-+}|^2+|m^+_{--}|^2)_{\chi_1\to 0}&=&\frac{4\vec{q}^{\,\,\,2}}{\chi_1}\frac{1}{x_1\chi_2\chi_2'}[(x')^2+(1-x_1)^2], \nonumber \\
(|m^+_{-+}|^2+|m^+_{--}|^2)_{\chi_1'\to 0}&=&\frac{4\vec{q}^{\,\,\,2}}{\chi_1'}\frac{(x')^3}{x_1(1-x_2)^2\chi_2\chi_2'}[1+(1-x_2)^2].
\ea

At small emission angles we can express all the invariants in terms of angular-two dimensional vectors in the
plane transversal to the beams axis:
\ba
\vec{k}_1=Ex_1\vec{\theta}_1,\quad \vec{p}_1'=Ex'\vec{\theta}',\quad
\int\limits_{\theta_1>\theta_0} \frac{d^2k_1}{\chi_1}=\pi x_1\ln\frac{1}{\theta_0^2}+...; \nonumber \\
\int\frac{d^2k_1}{\chi_1'}=\frac{x_1}{x'}\int\limits_{|\vec{\theta}_1-\vec{\theta}'|>\theta_0}\frac{d^2\theta_1}
{(\vec{\theta}_1-\vec{\theta}')^2}=\frac{\pi x_1}{x'}\ln\frac{1}{\theta_0^2}+... \!\!\!   .
\ea
One can be convinced explicitly in the absence of $\theta_0$-dependence in the sum of collinear kinematics
and the summed on final-state hard photon chiral states
of non-collinear contributions to electron impact factor:
\ba
\dd\tau^e_{hard,nc}=\sum_{\lambda_1\lambda_2}(\dd\tau^{e\gamma\gamma}_{+\lambda_1\lambda_2}+\dd\tau^{i,e}_{comp}+\dd\tau^{f,e}_{comp}).
\ea
Its value however depends essentially on the experimental photon detection set-up.
Similar calculations of  the photon impact factor in the non-collinear kinematics of the emission of hard photon
\begin{gather}
\gamma(k,\lambda_\gamma)+\gamma^*(q)\to e^-(q_-,\lambda_-)+e^+(q_+,-\lambda_+)+\gamma(k_1,\lambda_1),
\\ \nonumber
s_1=2q_-q_+,\quad \chi_\pm=2kq_\pm,\quad
\chi_{1\pm}=2k_1q_\pm,
\end{gather}
with chiral amplitudes defined as $m^{\lambda_\gamma}_{\lambda_1\lambda_-}$ gives
\ba
\dd\tau^{e^+e^-\gamma}_{\lambda_\gamma\lambda_1\lambda_-}
=\frac{\alpha^3}{2\pi^2}|m^{\lambda_\gamma}_{\lambda_1\lambda_-}|^2\frac{\dd^2q_-\dd^2q_+\dd x_+\dd x_-}{x_1x_+x_-}, \nonumber \\
x_1=1-x_+-x_-,\quad \vec{q}=\vec{q}_-+\vec{q}_++\vec{k}_1,
\ea
with
\begin{gather}
|m^+_{++}|^2=\frac{4\vec{q}^{\,\,\,2}s_1x_+^2}{\chi_-\chi_{1-}\chi_+\chi_{1+}},\quad
|m^+_{+-}|^2=\frac{4\vec{q}^{\,\,\,2}s_1x_-^2}{\chi_-\chi_{1-}\chi_+\chi_{1+}},\quad
|m^+_{--}(k,k_1)|^2=|m^+_{-+}(-k_1,-k)|^2,\nonumber \\
|m^+_{-+}|^2=\frac{4}{s_1^2\chi_-\chi_{1-}\chi_+\chi_{1+}}Tr\hat{q}_-A^+_{-+}\omega_+\hat{q}_+
\tilde{A}^+_{-+}.
\end{gather}
and
\ba
A^+_{-+}=\frac{s_1}{(q_+-q)^2}\hat{k}\hat{q}_+\hat{k}_1(-\hat{q}_++
\hat{q})\frac{\hat{p}_2}{s}-\frac{s_1}{(q_--q)^2}\frac{\hat{p}_2}{s}(\hat{q}_-
-\hat{q})\hat{k}\hat{q}_-\hat{k}_1 \nonumber \\
-\hat{q}_+(\hat{q}_--\hat{k})\frac{\hat{p}_2}{s}(\hat{q}_++\hat{k}_1)\hat{q}_-.
\ea
Again,  the proportionality $A^+_{-+}\sim|\vec{q}|$ at small $|\vec{q}|$ it was demonstrated in \cite{bbgk}.

To check the cancellation of the $\theta_0$ dependence, we put the limiting values of $|m^+_{\lambda_-\lambda_1}|^2$
in the limit of emission
angles close to the momentum directions of one of the charged particles.
\ba
(|m^+_{++}|^2+|m^+_{+-}|^2)_{\chi_{1-}\to 0}&=&\frac{4\vec{q}^{\,\,\,2}}{\chi_{1-}}\frac{(x_+)^2x_-}{x_1(1-x_+)^2\chi_+\chi_-}[x_-^2+(1-x_+)^2]; \nonumber \\
(|m^+_{-+}|^2+|m^+_{--}|^2)_{\chi_{1-}\to 0}&=&\frac{4\vec{q}^{\,\,\,2}}{\chi_{1-}}\frac{x_-}{x_1\chi_+\chi_-}[x_-^2+(1-x_+)^2]; \nonumber \\
(|m^+_{-+}|^2+|m^+_{--}|^2)_{\chi_{1+}\to 0}&=&\frac{4\vec{q}^{\,\,\,2}}{\chi_{1-}}\frac{x_+x_-}{x_1(1-x_-)^2\chi_+\chi_-}[x_+^2+(1-x_-)^2]; \nonumber \\
(|m^+_{++}|^2+|m^+_{+-}|^2)_{\chi_{1+}\to 0}&=&\frac{4\vec{q}^{\,\,\,2}}{\chi_{1-}}\frac{x_-}{x_1\chi_+\chi_-}[x_+^2+(1-x_-)^2].
\ea
One can be convinced in cancellation of $\theta_0$ dependence in the sum of collinear kinematics and the summed on
hard photon chiral states of non-collinear contributions to photon impact factor:
\ba
\dd\tau^\gamma_{hard,nc}=\dd\tau^{e^+e^-\gamma}_{+\lambda,\lambda}+\dd\tau^{-,\gamma}_{comp}+\dd\tau^{+,\gamma}_{comp}.
\ea
The numerical value of $\dd\tau^\gamma_{hard,nc}$ as well depends on experimental set-up and will not be considered here.

\section{Discussion and conclusions}

We have obtained that the impact factors of both electron and photon in LLA can be written in the partonic form of the
Drell-Yan process
in terms of structure functions for any chiral states of the initial and final particles
\ba
(\dd\tau_B+\dd\tau_{SV}+\sum\limits\dd\tau_{hard})^{\gamma,\,\lambda\sigma}(q_-,q_+)
=\int\limits_{x_-}^{1}\frac{\dd z_-}{z_-}\int\limits_{x_+}^1\frac{\dd z_+}{z_+}
D(\frac{x_-}{z_-},l_s)D(\frac{x_+}{z_+},l_s)\dd\tau^{\gamma,\,\lambda\sigma}_B(\frac{q_-}{z_-},\frac{q_+}{z_+}) \nonumber \\
\times(1+\frac{\alpha}{\pi}[K_{SV}^{\gamma,\,\lambda\sigma}+K^{-,\gamma}_{coll}
+K^{+,\gamma}_{coll}+K^\gamma_{ncol}]),
\nonumber \\
(\dd\tau_B+\dd\tau_{SV}+\sum\dd\tau_{hard})^{e,\,\sigma\lambda}(p_1,p_1')
=\int\limits_{0}^{1}\dd z_1\int\limits_{x'}^1\frac{\dd z_2}{z_2}
D(z_1,l_u)D(\frac{x'}{z_2},l_u)\dd\tau_B^{e,\,\sigma\lambda}(z_1p_1,\frac{p_1'}{z_2})\nonumber \\
\times(1+\frac{\alpha}{\pi}[K_{SV}^{e,\,\sigma\lambda}+K^{i,e}_{coll}+K^{f,e}_{coll}+K^e_{ncoll}]),
\nonumber
\ea
Here we suppress the chirality indices, and $D$ is nonsinglet structure function of fermion \cite{KF85}:
\begin{gather}
D(z,l)=\delta(z-1)+\frac{\alpha}{2\pi}(l-1)P^{(1)}(z)+\ldots \\ \nonumber
P^{(1)}(z)=\biggl(\delta(1-z)(2\ln\Delta+\frac{3}{2})+\theta(1-z-\Delta)\frac{1+z^2}{1-z}\biggr)_{\Delta\to 0}.
\label{kernelP}
\end{gather}

 The explicit form of $K_{SV}$ is presented in Appendix B  for the definite chiral states.
The explicit form of $K_{coll}$ is given above (see (\ref{kcoll_1},\ref{kcoll_2})). The form of $K_{ncol}$ (
after proper regularization-compensating the divergent terms in limit $\theta_0\to 0$) strongly depends on the details
of experiment-tagging the additional hard photon.
The nonleading terms are free from infrared and collinear divergences (
do not depend on $\lambda, \Delta$ and $\theta_0$).

The terms containing the vector product arise due to the nonzero imaginary part of the LP amplitudes.

We can be convinced in validity of gauge-invariance check: the squares of chiral amplitudes
in the Born approximation as well as one-loop corrected ones tend to zero as $|\vec{q}|^2$ at small $|\vec{q}|$.

The electron impact factor has also contributions from pair production channels \cite{bbgk} which
are not considered here.

The accuracy of the formulae given above are determined by the omitted terms (\ref{accur}):
\ba
1+\mathcal{O}\biggl(\frac{m^2}{s_1^2},\frac{m^2}{u^2},\biggl(\frac{\alpha}{\pi}\biggr)^2\biggr).
\ea

\begin{acknowledgments}
We are grateful to S. Bakmaev for attention to the work in the initial part of this paper.
One of us (EAK) is grateful to V. Serbo and V. Telnov for useful criticism and discussions.
\end{acknowledgments}

\section{Appendix A}
Here we put the asymptotic expressions for part of scalar,vector and tensor integrals corresponding to
the absorption of virtual photon by  electron from the pair created in $\gamma(p_1)\gamma^*(q)$ collisions.

We give first the scalar integrals with two, three, and four
(different) denominators
\begin{align}
(0)&=k^2-\lambda^2,\nn
(2)&=(q_--k)^2-m^2+\ii0=k^2-2q_-k+\ii0,\nn
(\bar{2})&=(-q_+-k)^2-m^2+\ii0,\nn
(q)&=(p_1-q_+-k)^2-m^2+\ii0.
\end{align}

The loop momentum
integrals with the denominator $(\bar{q})=(q_--p_1-k)^2-m^2$
instead of $(q)=(p_1-q_+-k)^2-m^2$, including scalar, vector and
tensor ones can be obtained from the ones listed below by means of
the replacement  (\ref{change}):
\begin{gather}
q_- \to -q_+,\quad q_+ \to -q_-,\quad p_1\to -p_1,\quad \chi_\pm
\to \chi_\mp,\quad (2)\to(\bar{2}),\quad (q)\to (\bar{q}).
\end{gather}
So we can restrict ourselves by consideration only the integrals
with denominators $(0)$, $(2)$, $(\bar{2})$, $(q)$.

In Appendix we use the same conservation law,
 on-mass shell conditions and  the kinematic invariants as for the photon impact factor (\ref{phot_proc},\ref{phot_desig})
\begin{equation}
s_1+\vec{q}^{\,\,\,2}=\chi_++\chi_-.
\end{equation}
Two denominator scalar integrals are defined as
$$I_{ij}=\int\frac{\dd^4k}{\ii\pi^2}\frac{1}{(i)(j)}\cdot$$
The explicit expressions for them are
\begin{align}
I_{02}&=L+1,& I_{2q}&=L-l_q+1,& I_{0q}&=L-l_++1, \nn
I_{0\bar{2}}&=L+1,& I_{2\bar{2}}&=L-L_s+1,& I_{\bar{2}q}&=L-1.
\end{align}
Here and below we use the notation
\begin{gather}
L=\ln\frac{\Lambda^2}{m^2},\quad
l_\pm=\ln\frac{\chi_\pm}{m^2},\quad l_q=\ln\frac{\vec{q}^{\,\,\,2}}{m^2},
\nn L_s=\ln\frac{s_1}{m^2}-\ii\pi=l_s-\ii\pi,\quad
l_l=\ln\frac{m^2}{\lambda^2}, \nn
\Li{2}{z}=-\int\limits_0^z\frac{\dd x}{x}\ln(1-x).
\end{gather}

Remind once more that we imply all the kinematic invariants to
be greater then electron mass squared $s_1\sim
\vec{q}^{\,\,\,2}\sim\chi_\pm\gg m^2$ and present below the asymptotic
expressions systematically omitting the terms of order $m^2/s_1$
and  similar ones.

The tree denominator scalar integrals
$I_{ijk}=\int\frac{\dd^4k}{\ii\pi^2(i)(j)(k)}$ are
\begin{align}
I_{0\bar{2}q}&=-\frac{1}{2\chi_+}\Big[l_+^2+\frac{2\pi^2}{3}\Big],
\qquad\qquad I_{02\bar{2}}=\frac{1}{2s_1}\Big[l_s^2+2l_sl_l-
\frac{4\pi^2}{3}-\ii\pi(2l_s+2l_l)\Big],& \nn
I_{2\bar{2}q}&=-\frac{1}{2(s_1+\vec{q}^{\,\,\,2})}\Big[l_q^2-l_s^2+\pi^2+2\ii\pi
l_s\Big],&\nn
I_{02q}&=\frac{1}{\chi_+-\vec{q}^{\,\,\,2}}\Big[l_q(l_q-l_+)+
\frac{1}{2}(l_q-l_+)^2+2\Li{2}{1-\frac{\chi_+}{\vec{q}^{\,\,\,2}}}\Big].&
\end{align}
The four denominator integral
$I_{02\bar{2}q}=\int\frac{\dd^4k}{\ii\pi^2(0)(2)(\bar{2})(q)}$
has the form
\begin{gather}
I_{02\bar{2}q}=\frac{1}{s_1\chi_+}\Big[l_q^2-2l_+l_s-l_sl_l+
2\Li{2}{1+\frac{\vec{q}^{\,\,\,2}}{s_1}}+\frac{\pi^2}{6}+
\ii\pi\big(2l_++l_l-2\ln(1+\frac{\vec{q}^{\,\,\,2}}{s_1})\big)\Big].
\end{gather}

Now we describe the vector integrals
\begin{equation} \label{eq:vecint}
I_{r}^{\mu}=\int\frac{\dd^4k k^\mu}{r}=
a^+_rq_+^\mu+a^-_rq_-^\mu+a^1_rp_1^\mu
\end{equation}
with $r=(ij),(ijk),(ijkl)$ where
$i,j,k,l=(0),(2),(\bar{2}),(q)$.

For the vector integrals with two denominators we have (we put
only nonzero coefficients)
\begin{align}
a^-_{2q}&=a^1_{2q}=-a^+_{2q}=\frac{1}{2}\Big(L-l_q+\frac{1}{2}\Big),&
&a^1_{0q}=-a^+_{0q}=\frac{1}{2}\Big(L-l_++\frac{1}{2}\Big),&\nn
a^-_{2\bar{2}}&=-a^+_{2\bar{2}}=\frac{1}{2}\Big(L-L_s+\frac{1}{2}\Big),&
&a^1_{\bar{2}q}=-\frac{1}{2}a^+_{\bar{2}q}=\frac{1}{2}\Big(L-\frac{3}{2}\Big),&\nn
a^-_{02}&=\frac{1}{2}L-\frac{1}{4},&
&a^+_{0\bar{2}}=-\frac{1}{2}L+\frac{1}{4}&
\end{align}
and the coefficients for the vector integrals with three
denominators are
\begin{align}
a^-_{02q}&=\frac{1}{a}\Big(\chi_+I_{02q}+\frac{2\chi_+}{a}l_+
-\frac{\vec{q}^{\,\,\,2}+\chi_+}{a}l_q\Big),&
&a^+_{02q}=-a^1_{02q}=\frac{1}{a}\Big(l_+-l_q\Big),& \nn
a^1_{0\bar{2}q}&=\frac{1}{\chi_+}\Big(-l_++2\Big),&
&a=\chi_+-\vec{q}^{\,\,\,2},& \nn
a^+_{0\bar{2}q}&=-I_{0\bar{2}q}-\frac{1}{\chi_+}l_+,&
&a^-_{02\bar{2}}=-a^+_{02\bar{2}}=\frac{1}{s_1}L_s,& \nn
a^-_{2\bar{2}q}&=\frac{1}{c}\Big(L_s-l_q\Big),&
&a^+_{2\bar{2}q}=-I_{2\bar{2}q}+\frac{1}{c}\Big(L_s-l_q\Big),& \nn
a^1_{2\bar{2}q}&=\frac{s_1}{c}I_{2\bar{2}q}+\frac{1}{c}\Big(-l_q+2\Big)
-\frac{2s_1}{c^2}\Big(L_s-l_q\Big),&
&c=s_1+\vec{q}^{\,\,\,2}=\chi_++\chi_-.&
\end{align}
Finally, the coefficient of the vector integral with 4
denominators has the form
\begin{align}
&a^1=\frac{s_1}{d}\Big(\chi_+A+\chi_-B-s_1C\Big),&
&a^+=\frac{\chi_-}{d}\Big(\chi_+A-\chi_-B+s_1C\Big)& \nn
&a^-=\frac{\chi_+}{d}\Big(-\chi_+A+\chi_-B+s_1C\Big),&
&d=2s_1\chi_+\chi_-,& \nn &A=I_{2\bar{2}q}-I_{0\bar{2}q},&
&B=I_{02q}-I_{2\bar{2}q},& \nn
&C=I_{02q}-I_{02\bar{2}}-\chi_+I_{02\bar{2}q}.& &&
\end{align}

We parameterized the second rank tensor integrals  in the form
\begin{multline}
I_{r}^{\mu\nu}=\int\frac{\dd^4k}{\ii\pi^2}\frac{k_\mu
k_\nu}{r}=\Big[a^g_r g+a^{11}_r
p_1p_1+a^{++}_rq_+q_++a^{--}_rq_-q_-+
a^{1+}_r(p_1q_++q_+p_1)+\\
a^{1-}_r(p_1q_-+q_-p_1)+a^{+-}_r(q_+q_-+q_-q_+)\Big]_{\mu\nu}.
\end{multline}
The coefficients for tensor integral with four denominators are
(we suppressed the index $02\bar{2}q$)
\begin{align}
&a^{1+}=\frac{1}{\chi_+}\Big(A_6+A_7-A_{10}\Big),&
&a^{+-}=\frac{1}{s_1}\Big(A_2+A_6-A_{10}\Big),&\nn
&a^{1-}=\frac{1}{\chi_-}\Big(A_2+A_7-A_{10}\Big),&
&a^{11}=\frac{1}{\chi_-}\Big(A_1-s_1a^{1+}\Big),& \nn
&a^{--}=\frac{1}{s_1}\Big(A_5-\chi_+a^{1-}\Big),&
&a^{++}=\frac{1}{s_1}\Big(A_3-\chi_-a^{1+}\Big),& \nn
&a^g=\frac{1}{2}\Big(A_{10}-A_2-\chi_+a^{1+}\Big),& &&
\end{align}
with
\begin{align}
&A_1=a^1_{2\bar{2}q}-a^1_{0\bar{2}q},&
&A_6=a^+_{02q}-a^+_{2\bar{2}q},& \nn
&A_2=a^-_{2\bar{2}q},&
&A_{7}=a^1_{02q}-\chi_+a^1,& \nn
&A_3=a^+_{2\bar{2}q}-a^+_{0\bar{2}q},&
&A_{8}=a^-_{02q}-a^-_{02\bar{2}}-\chi_+a^-,& \nn
&A_4=a^1_{02q}-a^1_{2\bar{2}q},&
&A_{9}=a^+_{02q}-a^+_{02\bar{2}}-\chi_+a^+,& \nn
&A_5=a^-_{02q}-a^-_{2\bar{2}q},& &A_{10}=I_{2\bar{2}q}.&
\label{eq:ejka}
\end{align}
One can verify that the checking relations
\begin{gather}
A_{4}=\chi_+a^{11}+s_1a^{1-},\quad
A_{8}=\chi_-a^{--}+\chi_+a^{+-},\quad
A_{9}=\chi_+a^{++}+\chi_-a^{+-},
\end{gather}
for the above coefficients (\ref{eq:ejka}) are fulfilled.

The coefficients entering into the tensor integral $I^{\mu\nu}_{02q}$ are
\begin{align}
&a^g_{02q}=\frac{1}{4}L+\frac{3}{8}+\frac{\vec{q}^{\,\,\,2}}{4a}l_q-
\frac{\chi_+}{a}l_+,& \nn &a^{+-}_{02q}=-a^{1-}_{02q}=\frac{1}{2a}
\Big[\frac{\chi_+}{a}(l_+-l_q)-1\Big],& \nn
&a^{++}_{02q}=a^{11}_{02q}=-a^{1+}_{02q}=\frac{1}{2a}(l_q-l_+),&
\nn
&a^{--}_{02q}=\frac{1}{a^2}\Big[\chi_+^2I_{02q}+\frac{3\chi_+^2}{a}l_+
+\frac{(\vec{q}^{\,\,\,2})^2-4\vec{q}^{\,\,\,2}\chi_+-3\chi^2_+}{2a}l_q+
\frac{\vec{q}^{\,\,\,2}-3\chi_+}{2}\Big].&
\end{align}
The coefficients entering into the tensor integral $I^{\mu\nu}_{02\bar{2}}$
are
\begin{gather}
a^g_{02\bar{2}}=\frac{1}{4}(L-L_s)+\frac{3}{8},\quad
a^{++}_{02\bar{2}}=a^{--}_{02\bar{2}}=\frac{1}{2s_1}(L_s-1),\quad
a^{+-}_{02\bar{2}}=-\frac{1}{2s_1},
\end{gather}
and the coefficients for the tensor integral $I^{\mu\nu}_{0\bar{2}q}$
are
\begin{align}
&a^g_{0\bar{2}q}=\frac{1}{4}(L-l_+)+\frac{3}{8},&
&a^{1+}_{0\bar{2}q}=\frac{1}{\chi_+}\Big(l_+-\frac{5}{2}\Big),&\nn
&a^{11}_{0\bar{2}q}=\frac{1}{2\chi_+}(-l_++2),&
&a^{++}_{0\bar{2}q}=I_{0\bar{2}q}+\frac{1}{2\chi_+}(3l_+-1).
\end{align}
In the case of the tensor integral $I^{\mu\nu}_{2\bar{2}q}$ they have
the form
\begin{align} \label{eq:a22q}
&a^g_{2\bar{2}q}=\frac{1}{2}\Big[\frac{1}{2}L+\frac{3}{4}-
\frac{s_1}{2c}L_s-\frac{\vec{q}^{\,\,\,2}}{2c}l_q \Big],\qquad
a^{--}_{2\bar{2}q}=-\frac{1}{2c}(l_q-L_s),& \nn
&a^{++}_{2\bar{2}q}=I_{2\bar{2}q}+\frac{3}{2c}(l_q-L_s),\quad\:\:\:\,
\qquad\qquad a^{+-}_{2\bar{2}q}=\frac{1}{2c}(l_q-L_s),& \nn
&a^{1-}_{2\bar{2}q}=\frac{1}{c}\Big[-\frac{1}{2}+\frac{s_1}{2c}L_s-
\frac{s_1}{2c}l_q\Big],& \nn
&a^{1+}_{2\bar{2}q}=\frac{1}{c}\Big[-\frac{5}{2}-s_1I_{2\bar{2}q}+
\frac{5s_1}{2c}L_s+\frac{2\vec{q}^{\,\,\,2}-3s_1}{2c}l_q\Big],& \nn
&a^{11}_{2\bar{2}q}=\frac{1}{c^2}\Big[4s_1+\vec{q}^{\,\,\,2}+
s_1^2I_{2\bar{2}q}-\frac{3s_1^2}{c}L_s+
\frac{3s_1^2-(\vec{q}^{\,\,\,2})^2-4s_1\vec{q}^{\,\,\,2}}{2c}l_q\Big].&
\end{align}
The checking equations for the coefficients (\ref{eq:a22q}) could
be obtained after multiplying $I^{\mu\nu}_{2\bar{2}q}$ by
$2(q_++q_-)^\nu$ or $2p_1^\nu$, using the relations
$2k\cdot(q_++q_-)=(\bar{2})-(2)$, $2p_1\cdot k
=(\bar{2})-(q)-\chi_+$ and using the vector integrals
(\ref{eq:vecint}). They have the form
\begin{align}
&2a^g_{2\bar{2}q}+s_1a^{--}_{2\bar{2}q}+ca^{1-}_{2\bar{2}q}+
s_1a^{+-}_{2\bar{2}q}=a_{2q}^--a_{\bar{2}q}^-,& \nn
&2a^g_{2\bar{2}q}+s_1a^{++}_{2\bar{2}q}+ca^{1+}_{2\bar{2}q}+
s_1a^{+-}=a_{2q}^+-a_{\bar{2}q}^+,& \nn
&ca^{11}_{2\bar{2}q}+s_1a^{1+}_{2\bar{2}q}+s_1a^{1-}_{2\bar{2}q}=
a_{2q}^1-a_{\bar{2}q}^1.&
\end{align}
Integrals for calculation of the electron impact factor with the denominators
\begin{align}
&(0)_e=k^2-\lambda^2,\nn
&(1)_e=(p_1 -k)^2-m^2+\ii0,\nn
&(2)_e=(p_1'-k)^2-m^2+\ii0,\nn
&(q)_e=(p_1-k_1-k)^2-m^2+\ii0.
\end{align}
can be obtained from the cited above by the substitution
\ba
\int\frac{d^4k}{i\pi^2}\;\frac{1,\,k,\,kk}{(0 \,1\,2\,q)_e}={\cal{P}}(q_- \to p_1',q_+ \to -p_1, p_1 \to -k_1, q \to q)
\int\frac{d^4k}{i\pi^2}\;\frac{1,\,k,\,kk}{(0 \,2\,\bar{2}\,q)}
\ea
An additional set of relevant integrals for the electron impact factor can be obtained
by the substitution (\ref{change_electron}).

\section{Appendix B}
The explicit expression for the photon K-factor are (the case of two different polarization):

\ba
\frac{1}{2}K^{\gamma +-}_{SV}=-\frac{1}{2}\ln^2 \frac{x_+}{x_-}-\frac{3}{4}l^2_{qs} +l_{qs}l_{ps}
+\frac{1}{2} l_{qs}l_{ms}+\frac{1}{4}l_{ms}+\frac{3}{4}l_{qs}+\frac{1}{2}l_{qs}-\frac{3}{4}
\nonumber  \\
+\frac{3}{2}\Li{2}{1+\frac{\vec{q}^{\,\,\,2}}{s_1}}-\Li{2}{ 1-\frac{\chi_+}{\vec{q}^{\,\,\,2}}}
-\frac{1}{2}\Li{2}{1-\frac{\chi_-}{\vec{q}^{\,\,\,2}}}
+x_+\frac{\chi_-x_+-2\chi_+-2s_1}{4x_-^2s_1}
 \nonumber \\
+\Big(\frac{ \chi_-x_+}{\chi_+x_-}-\frac{(x_+\chi_- -s_1)^2}{2x_-^2 \chi_+^2}\Big)
\Big( \frac{1}{2}l_{qs}^2-l_{qs}l_{ms} +\frac{\pi^2}{3} -\Li{2}{1+\frac{\vec{q}^{\,\,\,2}}{s_1}}+\Li{2}{1-\frac{\chi_-}{\vec{q}^{\,\,\,2}}}\Big)
\nonumber \\
-x_+\frac{(l_{ms}-l_{qs})}{x_-^2} \Big( \frac{a^2x_+ -2s_1 x_- \chi_-}{4 \tilde{a}^2} +\frac{4\chi_- -s_1 -2x_+ \chi_-}{2\tilde{a}}
- \frac{x_+\chi_-^2}{2\tilde{a} \chi_+} +\frac{2x_+ \chi_-- s_1}{2\chi_+x_+}   \Big)
\nonumber \\
+\frac{\pi}{2x_-^2\chi_+}[\vec{q}_+ \vec{q}_-]_z
\Big[\frac{l_{ms}-\ln\Big(1+\frac{\vec{q}^{\,\,\,2}}{s_1}\Big)}{\chi_+}(s_1-x_+\chi_-+x_-\chi_+)
 +\frac{s_1 +3x_-\chi_+-x_+\chi_-}{c} \Big]
\nonumber\\
+\frac{l_{qs}}{2x_-^2}\biggl(
\frac{ \chi_+^2+s_1^2+2x_+s_1\chi_+}{c^2}-\frac{2x_-s_1+\chi_+}{c}+\frac{s_1^2}{\chi_+c}
+x_+\frac{-2s_1+x_+\chi_-}{\chi_+}
\biggr)
\nonumber \\
+\frac{\chi_+^2+s_1^2+2x_+s_1\chi_+}{2x_-^2s_1c}
+\frac{\chi_+^2}{2s_1\vec{q}^{\,\,\,2}x_-^2}(\frac{1}{2}x_+^2-1)
\frac{\chi_+ x_+}{4\tilde{a}s_1\vec{q}^{\,\,\,2}x_-^2}\biggl(
x_+\chi_+^2+\vec{q}^{\,\,\,2}(x_+\chi_--2\chi_+)
\biggr)
\nonumber \\
+\frac{1}{\vec{q}^{\,\,\,2}x_-}\biggl(
\frac{1}{4}\chi_--\frac{\vec{q}^{\,\,\,2}}{2}+\chi_+x_--\frac{1}{2}s_1x_--\frac{1}{4}x_+^2s_1+\frac{1}{4}x_+^2\chi_+
\biggr)
 \;,
\ea

\ba
\frac{1}{2}K^{\gamma ++}_{SV}=-\frac{1}{2}\ln^2 \frac{x_+}{x_-}-\frac{3}{4}l_{qs}^2
+l_{qs}l_{ms} +\frac{1}{2}l_{qs}l_{ps}+\frac{1}{4}(l_{ps}+l_{qs})-\Li{2}{1-\frac{\chi_-}{\vec{q}^{\,\,\,2}}}
\nonumber \\
+\frac{3}{2}\Li{2}{1+\frac{\vec{q}^{\,\,\,2}}{s_1}}-\frac{1}{2}\Li{2}{ 1-\frac{\chi_+}{\vec{q}^{\,\,\,2}}}-1
\nonumber  \\
+\frac{2x_-x_+\chi_-\chi_+-(x_-\chi_+-s_1)^2}{2\chi_-^2x_+^2}
\Big( \frac{1}{2}l_{qs}^2-l_{qs}l_{ps}
-\Li{2}{1+\frac{\vec{q}^{\,\,\,2}}{s_1}}+\Li{2}{ 1-\frac{\chi_+}{\vec{q}^{\,\,\,2}}}+\frac{\pi^2}{3}  \Big)
 \nonumber \\
-\frac{l_{ps}-l_{qs}}{x_+^2} \Big[\frac{x_-(-2s_1\chi_+x_++x_-\tilde{a}^2)}{4a^2}
+\frac{x_-(4x_+s_1\chi_+-x_+s_1^2-x_-\tilde{a}^2)}{2as_1}
-\frac{(s_1-x_-\chi_+)^2}{2\chi_-s_1}  \Big]
\nonumber\\
+\frac{l_{qs}}{x_+^2}\Big[\frac{2x_+s_1\chi_++\tilde{a}^2}{2c^2}
+\frac{s_1^2}{2c\chi_-}-\frac{x_+\vec{q}^{\,\,\,2}+s_1x_-+\frac{1}{2}\chi_-}{c}
-\frac{x_-}{\chi_-}(s_1-\frac{1}{2}x_- \chi_+)-\frac{1}{2}-x_+x_-\Big]
\nonumber \\
+\pi \frac{[\vec{q}_+ \vec{q}_-]_z}{2x_+^2}
\Big[ \frac{x_+\chi_- + s_1-x_-\chi_+}{\chi_-^2} (l_{ps}-\ln\Big(1+\frac{\vec{q}^{\,\,\,2}}{s_1}\Big))
+\frac{2x_++1}{c}-\frac{x_-}{\chi_-}+\frac{s_1}{\chi_- c} \Big]
\nonumber \\
-\frac{1}{x_+^2}\biggl(
x_-\frac{2s_1-x_-\vec{q}^{\,\,\,2}}{\tilde{a}}+\frac{x_+x_-}{2}-\frac{2x_+s_1\chi_++(\chi_+-s_1)^2}{2cs_1}
\biggr)
\nonumber \\
+\frac{1}{\vec{q}^{\,\,\,2}x_+^2}\biggl(
\frac{x_-^2s_1^2}{4a}+\frac{\vec{q}^{\,\,\,2}}{4}+\frac{5}{4}\chi_++\frac{3}{2}s_1x_+-s_1-x_+\chi_+
-\frac{1}{2}x_+^2s_1
\biggr)
-\frac{\chi_+^2}{2s_1x_+^2\vec{q}^{\,\,\,2}},
\ea
\begin{gather}
 a=\chi_+-\vec{q}^{\,\,\,2},\quad \tilde{a}=\chi_- -\vec{q}^{\,\,\,2}, \quad c=\chi_++\chi_-,
\nonumber \\
l_{qs}=\ln\frac{\vec{q}^{\,\,\,2}}{s_1},\quad l_{ps}=\ln\frac{\chi_+}{s_1},\quad l_{ms}=\ln\frac{\chi_-}{s_1}
\end{gather}

The explicit expression for electron K-factor is:
\ba
K^{e,+-}_{SV}=-\ln^2x'-\frac{3}{2}\ln^2\frac{\vec{q}^{\,\,\,2}}{u}+2\ln\frac{\chi}{u}\ln\frac{\vec{q}^{\,\,\,2}}{u}
+\ln\frac{\chi'}{u}\ln\frac{\vec{q}^{\,\,\,2}}{u}+\frac{1}{2}\ln\frac{\chi'}{\vec{q}^{\,\,\,2}}
+3\ln\frac{\vec{q}^{\,\,\,2}}{u}
\nonumber \\
-2\Li{2}{1-\frac{\chi}{\vec{q}^{\,\,\,2}}}+3\Li{2}{1-\frac{\vec{q}^{\,\,\,2}}{u}}-\Li{2}{1+\frac{\chi'}{\vec{q}^{\,\,\,2}}}-\frac{\pi^2}{4}-2
\nonumber \\
-\frac{1}{(x')^2(\chi')^2}\biggl(
\frac{1}{2}\ln^2\frac{\vec{q}^{\,\,\,2}}{u}-\ln\frac{\vec{q}^{\,\,\,2}}{u}\ln\frac{\chi'}{u}-\Li{2}{1-\frac{\vec{q}^{\,\,\,2}}{u}}
+\Li{2}{1+\frac{\chi'}{\vec{q}^{\,\,\,2}}}+\frac{\pi^2}{12}
\biggr)
\nonumber \\
\times\biggl(d^2-2x'(-u\chi'+u^2+\chi\chi')+(x')^2u^2\biggr)
\nonumber \\
-\frac{1}{2(x')^2\vec{q}^{\,\,\,2}}(
-u+\chi+2x'\chi-(x')^2u-3(x')^2\chi+\frac{u^2}{\tilde{d}})
+\frac{(\chi')^2x^2}{(x')^2\vec{q}^{\,\,\,2}u}(1+\frac{\vec{q}^{\,\,\,2}}{c})
\nonumber \\
-\frac{1}{(x')^2}\biggl(
\frac{x(-u+2\chi+ux')}{c}+\frac{1}{2}-x'-\frac{\chi'-2u+2x'u}{2\tilde{d}}
\biggr)
\nonumber \\
-\frac{1}{(x')^2}\ln\frac{\chi'}{\vec{q}^{\,\,\,2}}\biggl(
\frac{x'(2d+x'u)}{\chi'}-\frac{x'(4\chi'-u)}{\tilde{d}}
-\frac{d^2(-2u-3\chi)}{2\tilde{d}^2\chi}
+\frac{x'u\chi'}{\tilde{d}^2}
\biggr)
\nonumber \\
-\frac{1}{(x')^2}\ln\frac{\vec{q}^{\,\,\,2}}{u}\biggl(
\frac{u\, x}{\chi(\chi')^2}
-\frac{x((\chi-u)^2x-2\chi u\, x')}{c^2}
-\frac{2u\, x\, x'}{c}+\frac{(u^2-\chi^2)x^2}{c\chi}
\biggr)
\nonumber \\
+4\pi\frac{[\vec{p}_1',\vec{k}_1]_z}{(x')^2}\biggl(\ln\frac{\vec{q}^{\,\,\,2}}{u}-\ln\frac{\chi'}{\vec{q}^{\,\,\,2}}\biggr)\frac{\chi'-u-x'(\chi-u)}{\chi^2}
+2\pi\frac{[\vec{p}_1',\vec{k}_1]_z}{(x')^2}\biggl(\frac{d(-2u-3\chi)}{\tilde{d}^2\chi}+x'\frac{-2u+\chi}{\chi\tilde{d}}\biggr),
\ea
where
\ba
d=\chi-\vec{q}^{\,\,\,2}, \quad \tilde{d}=u+\chi, \quad c=u-\vec{q}^{\,\,\,2},
\ea
For $K^{e,++}_{SV}$ we have:
\ba
K^{e,++}_{SV}=-\ln^2x'-\frac{3}{2}\ln^2\frac{\vec{q}^{\,\,\,2}}{u}+2\ln\frac{\vec{q}^{\,\,\,2}}{u}\ln\frac{\chi'}{u}
\ln\frac{\vec{q}^{\,\,\,2}}{u}+\ln\frac{\chi}{u}+\frac{1}{2}\ln\frac{\chi}{u}+\frac{5}{2}\ln\frac{\vec{q}^{\,\,\,2}}{u}
\nonumber \\
-2\Li{2}{1+\frac{\chi'}{\vec{q}^{\,\,\,2}}}+3\Li{2}{1-\frac{\vec{q}^{\,\,\,2}}{u}}-\Li{2}{1-\frac{\chi}{\vec{q}^{\,\,\,2}}}-\frac{\pi^2}{4}-
\frac{3}{2}
\nonumber \\
-\frac{1}{(\chi')^2}\biggl(-\ln^2\frac{\vec{q}^{\,\,\,2}}{u}+\ln\frac{\chi}{u}\ln\frac{\vec{q}^{\,\,\,2}}{u}-\frac{\pi^2}{12}
+\frac{(\chi')^2}{c^2}\ln\frac{\vec{q}^{\,\,\,2}}{u}+(\ln\frac{\chi}{u}-\ln\frac{\vec{q}^{\,\,\,2}}{u}
+\frac{\chi'}{c})\frac{\chi'}{u}
\nonumber \\
+\Li{2}{1-\frac{\vec{q}^{\,\,\,2}}{u}}-\Li{2}{1-\frac{\chi}{\vec{q}^{\,\,\,2}}}
\biggr)
\biggl(-u^2x^2+2x'\chi\tilde{d}+(x')^2\chi(-2u-\chi)\biggr)
\nonumber \\
+\frac{\chi^2(1-2x')}{u\, c}(1+\frac{u}{c}\ln\frac{\vec{q}^{\,\,\,2}}{u})
-\frac{2x'\chi}{u}\biggl(-\ln\frac{\chi}{u}-\frac{\chi'}{c}+\ln\frac{\vec{q}^{\,\,\,2}}{u}(1-\frac{u\chi'}{c^2})\biggr)
\nonumber \\
-\frac{x'}{d}(u\, x-\frac{1}{2}x'\chi)
+\frac{x'}{u\, d}\ln\frac{\chi}{\vec{q}^{\,\,\,2}}\biggl(u^2+4\chi u-x'\tilde{d}^2-\frac{u(-2\chi u +x' \tilde{d}^2)}{2d}\biggr)
\nonumber \\
-\ln\frac{\vec{q}^{\,\,\,2}}{u}\biggl(
x'\frac{-2ux+x'\chi}{\chi'}+\frac{(2u(1-(x')^2)-\chi x^2}{c}
-\frac{u^2x^2}{c\chi'}
\biggr)
\nonumber \\
-\frac{1}{\vec{q}^{\,\,\,2}}\biggl(
-\frac{1}{2}\chi+x'u+3x'\chi-2(x')^2u-\frac{5}{2}(x')^2\chi-\frac{d\, x^2 u^2}{2}
\biggr)
+\frac{x^2\chi^2}{u\vec{q}^{\,\,\,2}}.
\ea

\newpage

%

\end{document}